\documentclass{article}
\usepackage{arxiv}

\usepackage[utf8]{inputenc} 
\usepackage[T1]{fontenc}    
\usepackage[hidelinks]{hyperref}       
\usepackage{url}            
\usepackage{booktabs}       
\usepackage[version=4]{mhchem}
\usepackage{amsfonts}       
\usepackage{mathtools}
\usepackage{nicefrac}       
\usepackage{microtype}      
\usepackage{lipsum}		
\usepackage{graphicx}
\usepackage[square,numbers]{natbib}
\usepackage{doi}
\usepackage{caption}
\usepackage{subcaption}
\usepackage{amsmath}

\usepackage{booktabs}

\usepackage{bm}

\usepackage{url}
\usepackage[dvipsnames]{xcolor}
\definecolor{newcolor}{rgb}{.8,.349,.1}
\graphicspath{{Figure/}}

\newcommand{\Ma}{\operatorname{\mathit{M}}}

   \newcommand{\mean}[1]{\overline{#1}\,}
   
   \newcommand{\pmean}[1]{\overline{\overline{#1}}\,}
   \newcommand{\logmean}[1]{\overline{#1}^{\text{log}}}
   \newcommand{\logexpmean}[1]{\overline{#1}^{\text{log},\text{exp}}}
   
   \newcommand{\hmean}[1]{\overline{#1}^{H}}

\newcommand{\bfu}{\mathbf{u}}
\newcommand{\bff}{\mathbf{f}}

   \newcommand{\mF}{\mathcal{F}}

\newcommand{\dtp}{\delta^{\,+}}

\newcommand{\dd}{\mathrm{d}}

   \usepackage{caption}
   \usepackage{subcaption}
   \usepackage{epsfig,epstopdf}
   \usepackage[normalem]{ulem}
   
\title{Formulation of entropy-conservative discretizations for compressible flows of thermally perfect gases}%

\date{March 9, 2026}	

\author{ \href{https://orcid.org/0009-0003-6376-768X}{\includegraphics[scale=0.06]{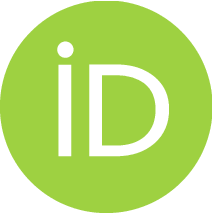}\hspace{1mm} Alessandro {Aiello}}\\
	Dipartimento di Ingegneria Industriale\\
	Universit\`a di Napoli ``Federico II''\\
	Napoli, Italy \\
	\texttt{alessandro.aiello@unina.it} \\
	\And
    \href{https://orcid.org/0000-0002-6518-3114}{\includegraphics[scale=0.06]{orcid.eps}\hspace{1mm} Carlo {De~Michele}}\\
	Gran Sasso Science Institute (GSSI)\\
	L'Aquila, Italy \\
	\texttt{carlo.demichele@gssi.it} \\
	\And
	\href{https://orcid.org/0000-0003-4943-9551}{\includegraphics[scale=0.06]{orcid.eps}\hspace{1mm}Gennaro Coppola} \\
	Dipartimento di Ingegneria Industriale\\
	Universit\`a di Napoli ``Federico II''\\
	Napoli, Italy \\
	\texttt{gcoppola@unina.it} \\
}

\renewcommand{\shorttitle}{
Formulation of EC discretizations for compressible flows of thermally perfect gases}%

\hypersetup{
pdftitle={Formulation of entropy-conservative discretizations for compressible flows of thermally perfect gases},
pdfsubject={physics.flu-dyn},
pdfauthor={Alessandro Aiello, Carlo De~Michele, Gennaro Coppola}
pdfkeywords={Compressible flows, Finite-volume, Entropy conservation, Thermally perfect gas, Kinetic-energy-preserving,},
}

\begin{document}
\maketitle

\begin{abstract}
This study proposes a novel spatial discretization procedure for the compressible Euler equations that guarantees entropy conservation at a discrete level
for thermally perfect gases. The procedure is based on a locally conservative formulation, and extends the entropy-conserving schemes to the more realistic case of thermally perfect gases, while still guaranteeing preservation of both linear invariants and kinetic energy. The proposed methodology, which can also be extended to multicomponent gases and to an Asymptotically Entropy-Conservative formulation, shows advantages in terms of accuracy and robustness when compared to existing similar approaches.
\end{abstract}


\section{Introduction} \label{sec:Introduction}
Accurate and robust numerical simulations play an invaluable role in the study of turbulent compressible flows; however, standard discretizations of the equations are known to suffer from nonlinear instabilities.
This is particularly pronounced at high Reynolds numbers, even in the absence of shock waves, and it is often attributed to the use of nondissipative schemes for the approximation of the convective terms~\cite{Coppola_AMR_2019}.

To address these challenges, research has been devoted to devising numerical discretizations capable of automatically reproducing important features of the continuous equations, such as enforcing physical symmetries or conserving primary or secondary invariants of the considered equations~\cite{Veldman_SIAMRev_2021}, or by imposing invariant domain constraints, such as positivity of the density or internal energy~\cite{Guermond_SIAM_2016,Guermond_SIAM_2018}.
These schemes, usually referred to as structure-preserving methods, have been shown to provide increased robustness and reliability, and have been the subject of many studies in recent years \cite{Verstappen_JCP_2003,Subbareddy_JCP_2009,Morinishi_JCP_2010,Capuano_JCP_2015a,Coppola_JCP_2019}.

In incompressible flows, kinetic energy acts as a bounded $L_2$ measure of velocity~\cite{Morinishi_JCP_1998}, and as such it acts as a mathematical entropy for the incompressible Euler equations and can be used to assess nonlinear stability.
In the context of compressible flow simulations, kinetic energy is no longer a conserved quantity and cannot provide a bound on the solution, but still the use of Kinetic Energy Preserving (KEP) methods has been successfully adapted from the incompressible case. 
These methods are the most established and widely used among structure-preserving schemes~\cite{Feiereisen_1981,Pirozzoli_JCP_2010,Coppola_JCP_2019,Edoh_JCP_2022}.
The conditions to obtain KEP schemes have been defined both in the context of finite volume methods~\cite{Jameson_JSC_2008b,Veldman_JCP_2019} and finite difference methods~\cite{Coppola_JCP_2019,Coppola_JCP_2023}.
However, the most commonly used definition for the KEP property 
introduces some indeterminacy, due to the arbitrariness of the discretization of the pressure term~
\cite{Ranocha_CAMC_2021,DeMichele_C&F_2023}, which nevertheless has been found to have a significant impact on the evolution of kinetic energy~\cite{Gassner_JCP_2016}.

In compressible flows, kinetic energy is not a mathematical entropy of the governing equations; instead, in the inviscid limit, this role is primarily assumed by the thermodynamic entropy, whose existence is closely linked to the symmetrization of the Euler equations~\cite{Friedrichs_PNAS_1971,Mock_JDE_1980}.
Accordingly, another important class of methods is that based on Entropy Conservative (EC) discretizations~\cite{Tadmor_AN_2003,Ismail_JCP_2009,Ranocha_JSC_2018}, which guarantee a correct induced discrete balance of entropy in absence of discontinuities. 
These methods have been typically used as baseline formulations to design entropy stable discretizations, which enforce semi-discrete entropy inequality~\cite{Tadmor_MC_1987}. This approach complements the more classical techniques based on artificial (or entropy) viscosity~\cite{Guermond_JCP_2011,Berthon_JSC_2023,Chan_JCP_2025,Kuzmin_CMAME_2022,Kurganov_JCP_2012}.
A fundamental difference between KEP and EC formulations is that, unlike KEP schemes, which only depend on the discretization of the continuity and momentum equations, EC schemes depend on the discretization of the energy equation, which, in turn, depends on the Equation of State (EoS) assumed for the gas.

In the case of a calorically perfect gas, several methods have been developed that can preserve the entropy balance at the discrete level both exactly \cite{Tadmor_AN_2003,Chandrashekar_CCP_2013,Ranocha_JSC_2018,DeMichele_JCP_2025} or approximately with arbitrary accuracy \cite{DeMichele_JCP_2023,Tamaki_JCP_2022,Kawai_JCP_2025}.
Most EC schemes have been developed within the framework of the technique introduced by \citet{Tadmor_MC_1987}, which has been applied to finite volume and finite element schemes.
An improvement in the affordability of EC schemes came from the work of 
\citet{Ismail_JCP_2009}, which introduced the use of fluxes based on the logarithmic mean serving as the foundation of many subsequent developments.
Building on this, \citet{Chandrashekar_CCP_2013} proposed the first scheme that simultaneously satisfied both the KEP and the EC property.
More recently, Ranocha~\cite{Ranocha_JSC_2018,Ranocha_2020,Ranocha_CAMC_2021} introduced a different KEP and EC formulation which, among other advantages, was shown to have a discretization of the pressure which led to more physically consistent treatment of kinetic energy~\cite{Gassner_JCP_2016}.

An alternative gas model is that of the thermally perfect gas, in which it is taken into account the temperature variation of specific heats
caused by the excitation of vibrational energy within the molecules of the gas and by the electronic energy associated with electron motion within the atoms and molecules~\cite{Anderson2019}.
The use of this model more accurately describes the behaviour at the high temperatures typical of combustion phenomena~\cite{Fedkiw_JCP_1997} as well as many wall-bounded high-speed flows, when temperature may exhibit both high absolute values and wide ranges \cite{Dong_2010,Jiang_ATE_2016,Jiang_AST_2019}.
In recent years, some extensions of the EC schemes to the case of thermally perfect gases have also been developed~\cite{Gouasmi_CMAME_2020,Peyvan_JCP_2023,Oblapenko_C&F_2025}. 
Notably, all these formulations reduce to the scheme of \citet{Chandrashekar_CCP_2013} in the limit of a calorically perfect gas, owing to an analogous treatment of the pressure terms.

In a recent paper~\cite{Aiello_JCP_2025}, the present authors developed a quite general methodology for the design of KEP and EC schemes for the system of compressible Euler equations with an arbitrary EoS, which can be reduced to the already known method of Ranocha~\cite{Ranocha_JSC_2018} in the case of a calorically perfect gas. Being completely general, the methodology can be easily applied to the case of a thermally perfect gas as well, which is a special case in which the perfect-gas EoS is still used in conjunction with an arbitrary relation between internal energy and absolute temperature only.
However, in its original formulation, the scheme exhibited a singularity in the limit of a constant temperature field, necessitating the local use of an alternative, nonsingular flux in such cases.

In this paper, we show how the peculiarities of the thermally perfect gas model can be exploited to particularize the general formulation developed for an arbitrary EoS, yielding EC discretizations suitable for smooth regions of the flow.
Specifically, this approach alleviates the intrinsic potential singularity of the original method.
Moreover, due to the difference in the treatment of the pressure terms, the formulation obtained is more robust and physically consistent as compared to existing ones.

The structure of the paper is as follows. In Section~\ref{sec:Problem_formulation}, we introduce the governing equations, including the thermodynamic closure for thermally perfect gases. We then recall the necessary conditions for kinetic energy preservation and entropy conservation, and show how the general EC condition particularizes to this case. A polynomial-based model for the temperature dependence of specific heats is adopted, leading to a discrete formulation that is EC, KEP, and essentially free of singularities. A comparison with existing schemes is also presented.

Section~\ref{sec:Generalization} extends the formulation to a different internal energy model, i.e.~the Rigid-Rotor Harmonic-Oscillator (RRHO) model;
we also generalize the scheme to multicomponent, nonreacting mixtures under a single temperature assumption.

Numerical results are provided in Section~\ref{Sec:Results} to assess the conservation properties and robustness of the schemes on two benchmark problems: an inviscid doubly periodic jet, and the Taylor--Green vortex.

Finally, Section~\ref{sec:Conclusions} presents concluding remarks, summarizes the findings, and outlines future research directions

\section{Problem formulation}
\label{sec:Problem_formulation}
The 1D compressible Euler equations can be written as 
\begin{equation}\label{eq:EulerEq}
    \dfrac{\partial \,\bfu}{\partial t} + \dfrac{\partial\, \bff (\bfu)}{\partial x} =0
\end{equation}
where $\bfu$ and $\bff$ are the variables and flux vectors, respectively. 
For the single-component case one has $\bfu = (\rho, \rho u, \rho E)^\text{T}$ and 
$\bff = (\rho u, \rho uu + p, u(\rho E + p))^\text{T}$ 
where $\rho$ and $\rho u$ are the density and momentum ($u$ being the velocity), $p$ is the pressure, and $E$ the total energy per unit mass, given by the sum of kinetic energy $\kappa = u^2/2$ and internal energy $e$: $E = \kappa + e$.
The one-dimensional case will be used for ease of exposition, but the theory presented here can be straightforwardly extended to the multidimensional case.
The thermally perfect gas model will be considered in this study, meaning that the usual perfect-gas EoS will be assumed: $p = \rho R T$, where $R$ is the gas constant and $T$ is the absolute temperature, but the isochoric specific heat capacity $c_v$ is dependent on temperature, which implies that internal energy can be expressed as 
\begin{equation*}
e = \int_{T_{\mathrm{ref}}}^{T} c_v(T')\, \dd{T'} + e_{\text{ref}}
\end{equation*}
and the ‘‘ref" subscript indicates some reference condition. 
Mayer's relation $c_p - c_v = R$ is still valid, $c_p$ being the isobaric specific heat.

The analysis is conducted within the framework of a conservative semidiscretized treatment of the governing equations, in which the time derivatives of the spatially discretized variables are driven by the difference of numerical fluxes at adjacent faces, which comprise convective and pressure contributions consistent with the components of $\bff$ in Eq.~\eqref{eq:EulerEq}. 
We work in a Finite Difference (FD) framework, although the numerical fluxes determined in Section~\ref{sec:Fluxes} and \ref{sec:RRHOmodel} can also be used with other discretization techniques (e.g.~with Finite Volume or Discontinuous Galerkin formulations). The method will be illustrated for second-order two-point fluxes on a uniform grid $x_i$ of width $h=x_{i+1}-x_i$; the corresponding high-order extension can be constructed through the approach used in \cite{DeMichele_JCP_2023,Aiello_JCP_2025,LeFloch_SIAMJNA_2002}. 
The numerical flux for a generic quantity $\rho \phi$ at the interface between cell $i$ and $i+1$ will be indicated by $\mF_{\rho \phi}$, whereas its convective contribution will be denoted by $\mF^\text{c}_{\rho \phi}$.
All the schemes we will consider in the analysis are KEP, which implies that the convective term in the momentum flux is given by the product between the mass flux $\mathcal{F}_{\rho}$ and the arithmetic average of velocity $\overline{u}=(u_{i} + u_{i+1})/2$.
This is indeed the necessary and sufficient condition for (second-order, two-point) fluxes to be KEP~\cite{Jameson_JSC_2008b,Veldman_JCP_2019}. 
An important assumption we make in this study is that the pressure term in the momentum equation is discretized using a second-order central derivative, giving rise to the general form for the momentum flux: $\mF_{\rho u} = \mF_{\rho}\mean{u} + \mean{p}$. As it is known, this formulation induces a locally conservative discretization of the convective term in the kinetic-energy equation, with flux $\mF^\text{c}_{\rho k} = \mF_{\rho} u_{i}u_{i+1}/2$ ~\cite{Coppola_JCP_2023}, and a central approximation of the non conservative pressure term in the form $u\partial p/\partial x \rightarrow u\delta p/\delta x$ with $\delta p = p_{i+1}-p_{i-1}$ and $\delta x = x_{i+1}-x_{i-1} = 2h$.

To be consistent with the induced evolution of kinetic energy, the total energy convective flux is designed as the sum of the kinetic and internal energy components $\mF^\text{c}_{\rho E}= \mF^\text{c}_{\rho\kappa}+\mF^\text{c}_{\rho e}$ and the conservative pressure term is also discretized to be compatible with a direct discretization of the internal-energy equation, which implies that the approximation of $\partial pu/\partial x$ in the total-energy equation is expressed by the advective form $p\delta u/\delta x + u\delta p/\delta x$, which is still conservative with numerical flux $\overline{\overline{\left(p,u\right)}}$, where $\overline{\overline{\left(\phi,\psi\right)}}=\left(\phi_i\psi_{i+1}+\phi_{i+1}\psi_i\right)/2$ is the product mean~\cite{Pirozzoli_JCP_2010,Coppola_JCP_2019}.
In conclusion, the general form of the KEP numerical fluxes we consider is
\begin{equation}\label{eq:General_Flux_Form}
\mathcal{F}_{\rho}= \mF^{\text{c}}_{\rho},\qquad\qquad
\mathcal{F}_{\rho u} =\mathcal{F}_{\rho}\,\overline{u}+\overline{p},\qquad\qquad
\mathcal{F}_{\rho E} = \mF^{\text{c}}_{\rho e} + \dfrac{1}{2}\mathcal{F}_{\rho}u_iu_{i+1}+\overline{\overline{\left(p,u\right)}}.
\end{equation}
In Eq.~\eqref{eq:General_Flux_Form} the convective fluxes for mass and internal energy are still unspecified, and are residual degrees of freedom at our disposal to enforce additional structural properties to the formulation. 
The objective of the present treatment is to derive a KEP discretization which is also conservative of entropy, defined for the thermally perfect gas as
\begin{equation*}
s = \int_{T_\mathrm{ref}}^{T} \frac{c_v(T')}{T'}\,\dd{T'} - R\log{(\rho/\rho_\mathrm{ref}}) + s_{\mathrm{ref}},
\end{equation*}
and with favourable properties in terms of robustness and accuracy.

\subsection{Derivation of the fluxes}\label{sec:Fluxes}
Our starting point is the formula expressing a general constraint for EC discretizations derived in \cite{Aiello_JCP_2025}, which is valid for an arbitrary equation of state:
\begin{equation}\label{eq:ECFlux_real}
    \mF^{\text{c}}_{\rho e} = \mF_{\rho}\,\dfrac{\dtp g/T}{\dtp 1/T } - \mean{u}\,\dfrac{\dtp p/T}{\dtp 1/T}.
\end{equation}
Eq.~\eqref{eq:ECFlux_real} makes use of the difference operator $\dtp \phi = \phi_{i+1} - \phi_{i}$ and introduces the Gibbs free energy $g = h- Ts$, with $h = e + p/\rho$ the  specific enthalpy.
Eq.~\eqref{eq:ECFlux_real} still does not constrain the mass flux $\mF_\rho$, leaving it free and giving rise to a family of EC discretizations, but it has the drawback of becoming singular in the case of constant temperature.
For thermally perfect gases one has $g = e+T(R-s)$, leading to
\begin{equation}\label{eq:gibbs_tp}
    g = \omega(T) + RT\log{\rho}; \quad\quad\quad \omega(T) = \int_{T_{\mathrm{ref}}}^{T}c_v(T')\,\dd{T'} - T \int_{T_\mathrm{ref}}^{T}\frac{c_v(T')}{T'}\,\dd{T'}  + T(R-R\log\rho_{\mathrm{ref}}-s_{\mathrm{ref}})+ e_{\mathrm{ref}},
\end{equation}
which separates the term with dependence on density from those dependent on temperature only.
By substituting Eq.~\eqref{eq:gibbs_tp} in Eq.~\eqref{eq:ECFlux_real} one has
\begin{equation*}\label{eq:ECFlux_real1}
    \mF^{\text{c}}_{\rho e} =\mF_{\rho}\,\dfrac{\dtp \omega/T}{\dtp 1/T } + \mF_{\rho}\,R\dfrac{\dtp \log \rho}{\dtp 1/T } - \mean{u}\,\dfrac{\dtp p/T}{\dtp 1/T}.
\end{equation*}
which, by using the perfect gas EoS can be rearranged as
\begin{equation}\label{eq:ECFlux_real2}
    \mF^{\text{c}}_{\rho e} =\mF_{\rho}\,\dfrac{\dtp \omega/T}{\dtp 1/T } + \,R\dfrac{\dtp \log \rho}{\dtp 1/T }\left(\mF_{\rho} - \mean{u}\,\dfrac{\dtp \rho}{\dtp \log \rho}\right).
\end{equation}
From Eq.~\eqref{eq:ECFlux_real2} 
it is possible to see that the dependence of the internal-energy numerical flux on pressure can be eliminated with a suitable choice of the mass flux.
In particular, the choice $\mF_{\rho}=\logmean{\rho}\mean{u}$, where $\logmean{\phi} = \dtp\phi/\dtp\log \phi$ is the logarithmic mean, nullifies the second term at the r.h.s. of Eq.~\eqref{eq:ECFlux_real2} and the final form for the convective flux of internal energy guaranteeing entropy conservation becomes
\begin{equation}\label{eq:ECFlux_tp}
    \mF^{\text{c}}_{\rho e} = \mF_{\rho}\,\dfrac{\dtp \omega/T}{\dtp 1/T } \qquad\text{with}\qquad
\mF_{\rho} = \logmean{\rho}\mean{u},
\end{equation}
where now the internal-energy flux is built as the product between the mass flux and a term constituting a consistent average of internal energy, which correctly depends only on temperature.
If the calorically perfect gas model is considered in place of the thermally perfect one, Eq.~\eqref{eq:ECFlux_tp} and \eqref{eq:General_Flux_Form} translate into the EC flux of \citet{Ranocha_JSC_2018} (see Eq.~\eqref{eq:Ranocha_Flux_Etot} below). 
For the thermally perfect gas, progress can be made only by assuming a specific functional dependence $c_v(T)$.

\subsection{Polynomial-based fitting model}\label{sec:PolynomialModel}

In combustion simulations, the polynomial-based approach is commonly used  when thermal equilibrium is assumed~\cite{Hansen_SANDIA_2019}, by which the isochoric specific heat is expressed using temperature-based polynomial fittings, although the final expression is not strictly a polynomial, since it can also include negative powers of temperature:
\begin{equation}\label{eq:cp_poly}
    {c}_v(T) = \sum_{m=-\ell}^r c_m {T}^m.
\end{equation}
In Eq.~\eqref{eq:cp_poly} $\ell$ and $r$ are arbitrary natural numbers, whose values are used to give flexibility to the fitting capability of the functional dependence $c_v(T)$.
We will detail the derivation of the EC fluxes for this general formulation, even though usually only 5, 7 or 9 coefficients are used to experimentally fit the gas behaviour~\cite{NASA_CHASE,NASA}. In addition to being widely used, this form of $c_v(T)$ is also easily tractable from an analytical point of view. For another example of functional dependence $c_v(T)$, see Section~\ref{sec:RRHOmodel}.

By substituting Eq.~\eqref{eq:cp_poly} into Eq.~\eqref{eq:gibbs_tp} and integrating we obtain
\begin{equation}
    \dfrac{\dtp \omega/T}{\dtp 1/T} =  \varepsilon_{\mathrm{ref}} + c_{-1}\left(1+\dfrac{\dtp\frac{\log{T}}{T}}{\dtp 1/T}\right) + c_0 \dfrac{\dtp\log{1/T}}{\dtp 1/T} -\sum_{\substack{m=-\ell\\m\ne -1,0}}^r\frac{c_m}{m(m+1)}\frac{\dtp T^{m}}{\dtp 1/T},  \label{eq:xi}
\end{equation}
where
\begin{equation*}
\varepsilon_\mathrm{ref} = e_\mathrm{ref}- c_{-1}\log{T_\mathrm{ref}}-\sum_{\substack{m=-\ell\\m\neq-1}}^r\frac{c_m}{m+1}T_{\mathrm{ref}}^{m+1}, 
\end{equation*}
which
is equivalent to the $b_1$ coefficient present in NASA polynomials in \cite{NASA}.
The variable term associated with $c_{-1}$ in Eq.~\eqref{eq:xi} can be manipulated as
\begin{equation*}
  \dfrac{\dtp\frac{\log{T}}{T}}{\dtp 1/T} = -\dfrac{\mean{1/T}}{\logmean{1/T}} + \mean{\log T},
\end{equation*}
while the terms involving $\dtp T^m$ can be simplified by using the general identities
\begin{equation}\label{eq:Diff_of_Powers}
    a^k-b^k = -\left(\frac{1}{a}-\frac{1}{b}\right)\sum_{\nu=1}^{k}a^{k-\nu+1}b^{\nu} =
     (a-b)\sum_{\nu=0}^{k-1}a^{k-\nu -1}b^\nu,
\end{equation}
which are valid for positive $k$.
In fact, from the first equality in Eq.~\eqref{eq:Diff_of_Powers} one has
\begin{equation*}
\dfrac{\dtp T^{m}}{\dtp 1/T}  = 
-\left(\sum_{\nu=1}^{m}T_{i+1}^{m-\nu+1}T_i^{\nu}\right) \qquad m\geq 1,
\end{equation*}
whereas the case of negative $m$ can be handled by using the second equality in Eq.~\eqref{eq:Diff_of_Powers} applied to the reciprocals, which leads to
\begin{equation*}
\dfrac{\dtp T^{m}}{\dtp 1/T}  = 
\left(\sum_{\nu=m+1}^{0}T_{i+1}^{m-\nu+1}T_i^{\nu}\right)\qquad m\leq-2.
\end{equation*}
We now define the averages
$$\overline{\overline{T}}^{\,m} = \dfrac{\sum_{\nu=1}^{m}T_{i+1}^{m-\nu+1}T_i^{\nu}}{m}\quad \text{for } m\geq 1\quad\quad\quad \text{and} \quad\quad\quad\widetilde{\widetilde{T}}^{\,m} = \dfrac{\sum_{\nu=m+1}^{0}T_{i+1}^{m-\nu+1}T_i^{\nu}}{|m|}\quad \text{for }m\leq -2,$$
both of which are consistent with $T^{m+1}$, with $\overline{\overline{T}}^{\,1} = \overline{\overline{\left(T,T\right)}}$ and 
$\widetilde{\widetilde{T}}^{\,-2} = \overline{\left(1/T\right)}$. 
The final result is
\begin{equation}\label{eq:ECFlux_exact}
    \mF^{\text{c}}_{\rho e} = \mF_{\rho}\left[e_{\mathrm{ref}} + 
    c_{-1}\left(1-\dfrac{\mean{1/T}}{\logmean{1/T}} + \mean{\log T}\right)+c_0\dfrac{1}{\logmean{1/T}} + \sum_{\substack{m=-\ell}}^{-2}\frac{c_m}{m+1}\widetilde{\widetilde{T}}^m +  
    \sum_{\substack{m=1}}^{r}\frac{c_m}{m+1}\overline{\overline{T}}^m 
    \right].
 \end{equation}
The scheme based on Eqs.~\eqref{eq:General_Flux_Form}, \eqref{eq:ECFlux_tp} and \eqref{eq:ECFlux_exact} is exactly entropy conserving for thermally perfect gases, and will be denoted as EC-TP. The use of the mass flux $\mF_{\rho}=\logmean{\rho}\mean{u}$ has reduced, in the case of thermally perfect gases, the singularity of the original formulation for an arbitrary EoS to that associated with the computation of the logarithmic mean, for which a local fix analogous to that devised by \citet{Ismail_JCP_2009} can be used. Asymptotically Entropy Conserving (AEC) formulations similar to that developed by \citet{DeMichele_JCP_2023} could also be easily derived by using the asymptotic expansion of the logarithm employed in the calorically perfect case. 
In fact, 
using the Taylor series expansion
\begin{equation*}
\dtp \log{\phi} =
 2\hat{\phi}\sum_{n=0}^\infty \frac{\hat{\phi}^{2n}}{2n+1},
 \label{eq:log_expansion}
\end{equation*} 
in which $\hat{\phi} =(\dtp \phi)/(2\mean{\phi}) $, and applying it to $\dtp \log{\rho}$ and $\dtp \log{T}$ in the logarithmic means $\logmean{\rho}$ and $\logmean{1/T}$ and truncating the sum to finite $N$, we obtain the class of AEC fluxes
\begin{equation}\label{eq:rhoFlux_AEC}
    \mF_{\rho}^{\text{AEC}(N)}= \mean{\rho}\mean{u}\left(\sum_{n=0}^N \frac{\hat{\rho}^{2n}}{2n+1}\right)^{-1},
\end{equation}
\begin{equation}\label{eq:ECFlux_AEC}
    \mF_{\rho e}^{\text{c AEC}(N)} = \mF_{\rho}^{\text{AEC}(N)}\left[\varepsilon_{\mathrm{ref}}+
    \left(c_0\hmean{T}-c_{-1}\right)
    \sum_{n=0}^N \frac{\hat{T}^{2n}}{2n+1}+
    c_{-1}\left(1+\mean{\log{T}}\right) + \sum_{\substack{m=-\ell}}^{-2}\frac{c_m}{m+1}\widetilde{\widetilde{T}}^m +  
    \sum_{\substack{m=1}}^{r}\frac{c_m}{m+1}\overline{\overline{T}}^m \right],
\end{equation}
where $\hmean{\phi}$ is the harmonic mean $\hmean{\phi}=\left(\mean{1/\phi}\right)^{-1}$.
As in the case of calorically perfect gases, this set of fluxes constitutes a nonsingular hierarchy of discretizations mainly based on simple algebraic operations and with increasing EC property, converging to Eq.~\eqref{eq:ECFlux_tp} and \eqref{eq:ECFlux_exact} as $N\to\infty.$

\subsection{Comparison with existing schemes}
In some applications, thermally perfect gases are studied using a strictly polynomial dependence for $c_v(T)$~\cite{Gouasmi_CMAME_2020}. In this case, negative indices $m$ in Eq.~\eqref{eq:cp_poly} and \eqref{eq:ECFlux_exact} can be eliminated, and the resulting set of EC fluxes for thermally perfect gases reads
\begin{equation}\label{eq:ECFlux_exact_pos}
\mathcal{F}_{\rho}= \logmean{\rho}\mean{u}\qquad
\mathcal{F}_{\rho u} =\mathcal{F}_{\rho}\,\overline{u}+\overline{p},\qquad
\mathcal{F}_{\rho E} =\mF_{\rho}\left[\varepsilon_{\mathrm{ref}} + 
c_0\dfrac{1}{\logmean{1/T}} + 
    \sum_{\substack{m=1}}^{r}\frac{c_m}{m+1}\overline{\overline{T}}^m 
     + \dfrac{u_iu_{i+1}}{2}\right]+\overline{\overline{\left(p,u\right)}}.
\end{equation}
This set of fluxes can be compared to the EC fluxes derived by \citet{Gouasmi_CMAME_2020}, who analysed the same problem with a polynomial dependence for $c_v(T)$ by using the theory by Tadmor~\cite{Tadmor_AN_2003,Tadmor_MC_1987}.
In the case of single-component gas, their formula reads, in our notation\footnote{It should be noted that, in the original formulation by \citet{Gouasmi_CMAME_2020}, $e_{\mathrm{ref}}$ was used in place of $\varepsilon_{\mathrm{ref}}$ since it was assumed that $T_{\mathrm{ref}}$ would be equal to zero.},
\begin{equation}\label{eq:ECFlux_Gouasmi}
\mathcal{F}_{\rho}= \logmean{\rho}\mean{u}\qquad
\mathcal{F}_{\rho u} =\mathcal{F}_{\rho}\,\overline{u}+
\textcolor{red}{\dfrac{R\,\mean{\rho}}{\mean{1/T}}},\qquad
\mathcal{F}_{\rho E} =\mF_{\rho}\left[\varepsilon_{\mathrm{ref}} + 
c_0\dfrac{1}{\logmean{1/T}} + 
    \sum_{\substack{m=1}}^{r}\frac{c_m}{m+1}\overline{\overline{T}}^m 
     + \dfrac{u_iu_{i+1}}{2}\right]+\textcolor{red}{\mean{u}\dfrac{R\,\mean{\rho}}{\mean{1/T}}}.
\end{equation}
By comparing Eqs.~\eqref{eq:ECFlux_exact_pos} and \eqref{eq:ECFlux_Gouasmi} one observes that the two sets of fluxes are almost coincident, except for the pressure terms in the momentum and total-energy equations, highlighted in red in Eq.~\eqref{eq:ECFlux_Gouasmi}. While in the present formulation (Eq.~\eqref{eq:ECFlux_exact_pos}) the pressure flux in the momentum equation is expressed through the arithmetic mean $\mean{p}$, in the formulation by \citeauthor{Gouasmi_CMAME_2020} (Eq.~\eqref{eq:ECFlux_Gouasmi}) it is expressed by using the complex average $\widehat{p} = R\,\mean{\rho}/\mean{1/T}$, which depends on density and temperature. Moreover, the pressure term in the total-energy equation is expressed as $\pmean{(p,u)}$ in Eq.~\eqref{eq:ECFlux_exact_pos} and as $\widehat{p}\,\,\mean{u}$ in Eq.~\eqref{eq:ECFlux_Gouasmi}. 
It is well known that the discretization of the pressure term in the momentum equation plays an important role in the global discrete preservation of kinetic energy. In fact, in \cite{Gassner_JCP_2016}, which deals with KEP split forms for calorically perfect gases, it is shown that even in the case in which the convective flux in the momentum equation is written in the KEP form as $\mF^{\text{c}}_{\rho u} = \mF_{\rho}\mean{u}$, the discretization of the pressure term can spoil the correct balance of induced kinetic energy causing errors in the pressure work contribution.
This is especially true when, as in the case of the Chandrashekar scheme \cite{Chandrashekar_CCP_2013} to which Eq.~\eqref{eq:ECFlux_Gouasmi} reduces for ideal gases, the pressure average is rendered as a combination of averages of density and temperature.
The conclusions of the analysis conducted in~\cite{Gassner_JCP_2016} are that, in the case of calorically perfect gases, the simple arithmetic average of pressure in the momentum flux is the choice yielding the best results for the case of inviscid Taylor--Green vortices. In Section~\ref{Sec:Results_TGV} we show that this result also applies in the case of thermally perfect gases.
Similarly, the treatment of the pressure term in the total energy equation is known to be an element potentially affecting certain statistical quantities related to the fluctuating motion in homogeneous isotropic turbulence. In \cite{DeMichele_C&F_2023} and \cite{Pirozzoli_JCP_2010}, as an example, it is shown that r.m.s of density and temperature fluctuations in homogeneous isotropic turbulence simulations can erroneously grow without bound if the pressure term in the total energy equation is different from $\pmean{(p,u)}$. 
In Section~\ref{Sec:Results_TGV} we show that  this is the case also for thermally perfect gases, demonstrating the importance of the set of fluxes here derived as a relevant improvement on existing formulations.

\section{Generalizations and extensions}
\label{sec:Generalization}
\subsection{Numerical fluxes for hypersonic flows with the RRHO model}\label{sec:RRHOmodel}
The general methodology illustrated in Section~\ref{sec:Fluxes} can be used to derive EC numerical fluxes for arbitrary thermally perfect models in which the constant-volume specific heat $c_v(T)$ is amenable to an analytical treatment. 
In this section, we give a further example of the procedure by analyzing the Rigid-Rotor Harmonic-Oscillator (RRHO) model used for hypersonic flows' simulations as reported by \cite{Peyvan_JCP_2023} and adapted from a formulation presented in~\cite{Marxen_JCP_2013}.
The model takes into account vibrational and rotational contributions to internal energy, which for a single diatomic gas takes the form 
\begin{equation}\label{eq:RRHO_IntEnergy}
    e(T) = e_{\mathrm{ref}} +aRT+\dfrac{R\theta}{e^{\frac{\theta}{T}}-1},
\end{equation}
where $a$ equals $5/2$ for diatomic molecules and $\theta$ is the vibrational characteristic temperature associated with the single vibrational mode.
The constant-volume specific heat and the entropy $s$ are easily obtained from Eq.~\eqref{eq:RRHO_IntEnergy} and read
\begin{equation*}
    c_v(T) = aR + R\dfrac{\frac{\theta^2}{T^2}e^{\frac{\theta}{T}}}{\left(e^{\frac{\theta}{T}}-1\right)^2},\qquad
    s = s_{\mathrm{ref}} + R\left(a\log T +\dfrac{\theta}{T} + \dfrac{\frac{\theta}{T}}{e^{\frac{\theta}{T}}-1}-\log(e^{\frac{\theta}{T}}-1)-\log\rho\right) + \text{const.},
\end{equation*}
from which the Gibbs free energy results
\begin{equation*}
    g(T) = \underbrace{e_{\mathrm{ref}} - RT\left(a\log T - \log(e^{\frac{\theta}{T}}-1)+\dfrac{\theta}{T} + \text{const.}\right)}_{\omega(T)} + RT\log\rho.
\end{equation*}
By using Eq.~\eqref{eq:ECFlux_tp}, straightforward manipulations eventually give
\begin{equation*}
\mF^c_{\rho e} = \logmean{\rho}\mean{u}\left(e_{\mathrm{ref}} -R\theta + \dfrac{aR}{\logmean{1/T}} + \dfrac{R}{\logexpmean{1/T}}
\right),
\end{equation*}
where in analogy with~\cite{Peyvan_JCP_2023} we define the new \emph{log-exp} mean as
$\logexpmean{\phi} = \delta^+\phi/\delta^+\log\left(e^{\theta\phi}-1\right)$.
The final form of the fluxes is
\begin{equation}\label{eq:ECFlux_RRHO}
\mathcal{F}_{\rho}= \logmean{\rho}\mean{u}\qquad
\mathcal{F}_{\rho u} =\mathcal{F}_{\rho}\,\overline{u}+\overline{p},\qquad
\mathcal{F}_{\rho E} =\mF_{\rho}\left[e_{\mathrm{ref}} -R\theta + \dfrac{aR}{\logmean{1/T}} + \dfrac{R}{\logexpmean{1/T}}
     + \dfrac{u_iu_{i+1}}{2}\right]+\overline{\overline{\left(p,u\right)}},
\end{equation}
which is again coincident with the single component version of Eqs.~(63)--(65) in~\cite{Peyvan_JCP_2023}, except for the pressure terms in momentum and total-energy equations, which are there discretized as in Eq.~\eqref{eq:ECFlux_Gouasmi}.

\subsection{Multicomponent flows}
In the case of a gas composed of $N_s$ nonreacting species, the usual standard approach considers the mixture as a single compressible fluid, with density, momentum and total energy (per unit volume) given by the sum of the individual components, where the density of each species evolves according to a separate continuity equation~\cite{Abgrall_JCP_1996,Gouasmi_CMAME_2020,Renac_JCP_2021,Fujiwara_JCP_2023}. In this case, the compressible Euler equations are still in the form of Eq.~\eqref{eq:EulerEq} where $\bfu = \left(\rho_1, \rho_2,\ldots,\rho_N,\rho u,\rho E\right)^{\text{T}}$ and $\bff = \left(\rho_1u, \rho_2u,\ldots,\rho_{N_s}u,\rho uu +p,u(\rho E + p)\right)^{\text{T}}$, 
where $\rho_k = \rho Y_k$ is the density of the $k$-th species, $Y_k$ being the mass fraction and $\rho$ the density of the mixture. From the relation $\sum_k Y_k=1$ it follows $\sum_k\rho_k=\sum_k\rho Y_k=\rho$.
Internal energy and entropy of the mixture are defined as $\rho e=\sum_k\rho_k e_k$ and $\rho s=\sum_k\rho_k s_k$, which also implies $e=\sum_k Y_k e_k$ and $s=\sum_k Y_k s_k$. Individual equations of state $p_k = \rho_k R_k T$, where $R_k$ is the gas constant of the $k$-th species, and Dalton's law $\sum_kp_k = p$ are assumed.
The Gibbs' relation, which holds for each species: $d\left(\rho_ke_k\right) = Td\left(\rho_ks_k\right)+g_kd\rho_k$, 
can be used to relate the convective and pressure terms in the balance equations for mass, internal energy and entropy for each component of the mixture, as it was done for a single compound in \cite{Aiello_JCP_2025}.
By imposing a locally conservative formulation for mass and internal energy, and a straightforward discretization of the pressure term, one obtains the analogous sufficient conditions for entropy conservation of each species:
\begin{equation}\label{eq:ECFlux_real_k}
    \mF^{\text{c}}_{\rho_k e_k} = \mF_{\rho_k}\,\dfrac{\dtp g_k/T}{\dtp 1/T } - \mean{u}\,\dfrac{\dtp p_k/T}{\dtp 1/T},\qquad\qquad k=1\ldots N_s
\end{equation}
where the internal-energy flux of the mixture is $\mF^{\text{c}}_{\rho e} = \sum_k\mF^{\text{c}}_{\rho_k e_k}$.
Assuming the thermally perfect gas model and $\mF_{\rho_k} = \logmean{\rho_k}\mean{u}$
one obtains
\begin{equation}\label{eq:ECFlux_tp_k}
    \mF^{\text{c}}_{\rho_k e_k} = \mF_{\rho_k}\,\dfrac{\dtp \omega_k/T}{\dtp 1/T },
\end{equation}
with $g_k = \omega_k(T) + R_kT\log{\rho_k}$ 
as in Eq.~\eqref{eq:gibbs_tp}, where
internal energy and entropy for each thermally perfect component are given by $e_k = \int_{T_{\mathrm{ref}}}^{T} c_{v\,k}(T')\, \dd{T'} + e_{k\,\text{ref}}$ and $s_k = \int_{T_\mathrm{ref}}^{T} \frac{c_{v\,k}(T')}{T'}\,\dd{T'} - R_k\log{(\rho_k/\rho_{k\,\mathrm{ref}}}) + s_{k\,\mathrm{ref}}$.
Eq.~\eqref{eq:ECFlux_tp_k} is the general sufficient condition for entropy preservation of each species in the mixture.
 
In the case of polynomial approximation for $c_{v\,k}(T)= \sum_{m=0}^r c_{m}^k {T}^m $ the final set of fluxes reads
\begin{equation}\label{eq:ECFlux_exact_k}
\mathcal{F}_{\rho_k}= \logmean{\rho_k}\mean{u}\qquad
\mathcal{F}_{\rho u} =\sum_{k=1}^{N_s}\mathcal{F}_{\rho_k}\,\overline{u}+\overline{p},\qquad
\mathcal{F}_{\rho E} =\sum_{k=1}^{N_s}\mF_{\rho_k}\left[\varepsilon_{k\,\mathrm{ref}} + 
c_{0}^k\dfrac{1}{\logmean{1/T}} + \sum_{\substack{m=1}}^{r}\frac{c_{m}^k}{m+1}\overline{\overline{T}}^m 
     + \dfrac{u_iu_{i+1}}{2}\right]+\overline{\overline{\left(p,u\right)}}.
\end{equation}
which can be compared to Eq.~(44) of~\cite{Gouasmi_CMAME_2020}, where the same considerations on the discretization of the pressure terms still hold in this case.

For calorically perfect gases, the fluxes reduce to
\begin{equation}\label{eq:ECFlux_exact_CP_k}
\mathcal{F}_{\rho_k}= \logmean{\rho_k}\mean{u}\qquad
\mathcal{F}_{\rho u} =\sum_{k=1}^{N_s}\mathcal{F}_{\rho_k}\,\overline{u}+\overline{p},\qquad
\mathcal{F}_{\rho E} =\sum_{k=1}^{N_s}\mF_{\rho_k}\left[\varepsilon_{k\,\mathrm{ref}} +
\left(\logmean{1/e_k}\right)^{-1} 
     + \dfrac{u_iu_{i+1}}{2}\right]+\overline{\overline{\left(p,u\right)}},
\end{equation}
which is a possible  extension of the Ranocha's flux to multicomponent flows of ideal gases.

\section{Numerical results} \label{Sec:Results}

Numerical tests are presented to assess the EC properties and the robustness of the proposed formulation; all simulations have been carried out using the dimensionless form of the equations. Time integration has been performed with a standard fourth-order explicit Runge--Kutta method, with a sufficiently small CFL number to limit time-integration errors.
For the thermodynamic framework, we have adopted the experimental fit for $c_v(T)$ given by the five-coefficient approximation reported in \cite{NASA_CHASE}, where the only nonzero coefficients are $c_{-2},\, c_0,\, c_1,\, c_2,\, c_3$, whose values are reported in Tab.~\ref{tab:coeff} for convenience.
\begin{table}[h!]
\centering
\caption{NASA 5-coefficient polynomial for CH$_4$ (Low temperature range), \cite{NASA_CHASE}.}
\begin{tabular}{ll}
\toprule
Temperature range (K) & $298-1300$ \\
\midrule
$c_{-2}$ & 0.678565 \\
$c_0$ & -0.703029		 \\
$c_1$ & 108.4773	 \\
$c_2$ &-42.52157	 \\
$c_3$ &5.862788		 \\
\bottomrule
\end{tabular}
\label{tab:coeff}
\end{table}

In the dimensionless framework $R=1$, thus the equation of state reads $p=\rho T$ and the Mayer's relation is $c_p-c_v=1$. Reference energy, which is used to nondimensionalize internal energy as well as enthalpy, is $E_{\mathrm{r}}^*=R^*T_{\mathrm{r}}^*$, with the asterisk indicating dimensional quantities.
In the following results, the specific gas chosen is high-enthalpy methane, whose heat capacities strongly depend on temperature~\cite{NASA_CHASE,NASA}. 
Moreover, the nonpolarity of such molecule allows van der Waals forces to be neglected, especially at high temperatures where its modelling as a thermally perfect gas is preferable. 
Regarding the spatial discretization, we use the novel formulation~\eqref{eq:rhoFlux_AEC}--\eqref{eq:ECFlux_AEC} (together with the numerical flux for momentum given in Eq.~\eqref{eq:General_Flux_Form}), here termed AEC-TP$^{(N)}$, with a sufficiently high value of $N$; in all our tests we 
checked that the value of $N$ guaranteeing the convergence to the EC scheme in Eq.~\eqref{eq:ECFlux_exact} is typically not greater than $5$.
The novel discretization is compared to the one proposed by \citeauthor{Gouasmi_CMAME_2020}, reported in Eq.~\eqref{eq:ECFlux_Gouasmi} and generalized to include also negative exponents in the expression of the specific heat capacities.
Additionally, results are provided for other widely used discretizations, such as the formulation proposed by Ranocha~\cite{Ranocha_JSC_2018}
\begin{equation}\label{eq:Ranocha_Flux_Etot}
\mathcal{F}_{\rho}= \logmean{\rho}\mean{u},\qquad\qquad
\mathcal{F}_{\rho u} =\mathcal{F}_{\rho}\,\mean{u} + \mean{p},\qquad\qquad
\mathcal{F}_{\rho E} = \mathcal{F}_{\rho}\,\left[\left(\logmean{1/e}\right)^{-1} +\frac{u_iu_{i+1}}{2}\right] + \pmean{(p,u)},
\end{equation}
the KEEP formulation proposed by \citet{Kuya_JCP_2018}, whose fluxes are
\begin{equation}\label{eq:KEEP_Flux_Etot}
\mathcal{F}_{\rho}= \mean{\rho}\mean{u},\qquad\qquad
\mathcal{F}_{\rho u} =\mathcal{F}_{\rho}\,\mean{u} + \mean{p},\qquad\qquad
\mathcal{F}_{\rho E} = \mathcal{F}_{\rho}\left[\mean{e} + \frac{u_iu_{i+1}}{2}\right] + \pmean{(p,u)},
\end{equation}
and the splitting proposed by Jameson and Pirozzoli~\cite{Jameson_JSC_2008b, Pirozzoli_JCP_2010} (hereunder denoted as JP) whose expression is
\begin{equation}\label{eq:Flux_JP}
\mathcal{F}_{\rho}= \mean{\rho}\mean{u},\qquad\qquad
\mathcal{F}_{\rho u} =\mathcal{F}_{\rho}\,\mean{u} + \mean{p},\qquad\qquad
\mathcal{F}_{\rho E} = \mF_{\rho}\left[\mean{e}+\mean{\frac{u^2}{2}}\right] + \mF_{\rho}\mean{\left(\frac{p}{\rho}\right)}.
\end{equation}
The Ranocha's flux in Eq.~\eqref{eq:Ranocha_Flux_Etot} is exactly EC for calorically perfect gases, and is also KEP. The KEEP and JP fluxes in Eq.~\eqref{eq:KEEP_Flux_Etot} and \eqref{eq:Flux_JP} are KEP, but not EC, although the KEEP formulation has noticeable entropy conservation properties for calorically perfect gases~\cite{DeMichele_C&F_2023,Tamaki_JCP_2022,DeMichele_JCP_2023}.
In order to better evaluate the effect that the discretizations have on the evolution of quantities such as entropy and kinetic energy, we define a normalized difference of the integral value with respect to its value at time zero. For a generic variable $\phi$, that is
\begin{equation*}
    \langle \rho \phi \rangle =
\dfrac{\int_{\Omega}  \rho \phi \,\dd \Omega - \int_{\Omega} \rho_0 \phi_0\, \dd \Omega}{ \left\lvert\int_{\Omega} \rho_0 \phi_0 \,\dd \Omega\right\rvert}.
 \end{equation*}
\subsection{Inviscid doubly periodic jet flow} \label{Sec:Results_Jet}
In this section we report the case of the inviscid doubly periodic jet flow. The flow is solved in the periodic domain $(x,y)\in[0,L]\times[-L/4,L/4]$, with $L=1$, on a coarse mesh consisting of $N_P=65\times33$ evenly spaced points by means of second-order, central, spatial discretizations. CFL is set equal to $0.01$.
The initial conditions are chosen as follows:
\begin{equation}
    \begin{dcases}
        u(x,y^\pm,0) & = A_u\left[1 \mp  B_u\tanh{\dfrac{y\mp L/10}{\delta}}\right]\\
        v(x,y,0) &=\epsilon\sin{(2n\pi x/L)}\\        
        T(x,y^\pm,0) & =A_T\left[\dfrac{3}{2} \pm B_T\tanh{\dfrac{y\mp L/10}{\delta}}\right]\\
        p(x,y,0) &=p_0
    \end{dcases}
\end{equation}
where we set $A_u=B_u=B_T =1/2,\, A_T=2$ and $\delta=1/25$. For the y-velocity component, we have  
$\epsilon=0.01$ and $n=3$ to trigger the formation of three roll-up vortices. The characteristic time for this flow is defined as $t_{\mathrm{c}}=n^{-1}/\max_{x,y}u(x,y,0) \approx 0.44 $. A bulk pressure value of $p_0=2$ is assigned and the resulting density is $\rho=p/RT$.

The double-jet flow has been chosen because it displays a wide temperature range, which makes the hypothesis of calorically perfect gas too stringent---especially for chemical species such as methane.
Second-order spatial discretization is used here to test the basic versions of the fluxes as discussed in Section~\ref{sec:Fluxes}; high-order formulations are used in the next test case in Section~\ref{Sec:Results_TGV}.

\begin{figure}[tb]
    \centering
     \subfloat[Entropy]{\includegraphics[width=0.5\linewidth]{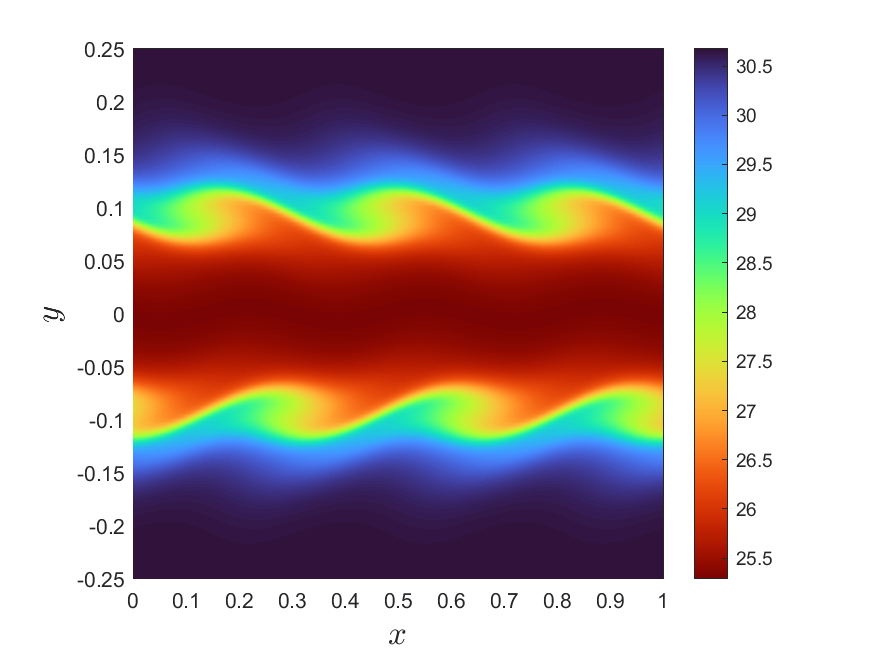}\label{fig:INSTDJ_s}}
     \subfloat[Thermodynamic pressure]{\includegraphics[width=0.5\linewidth]{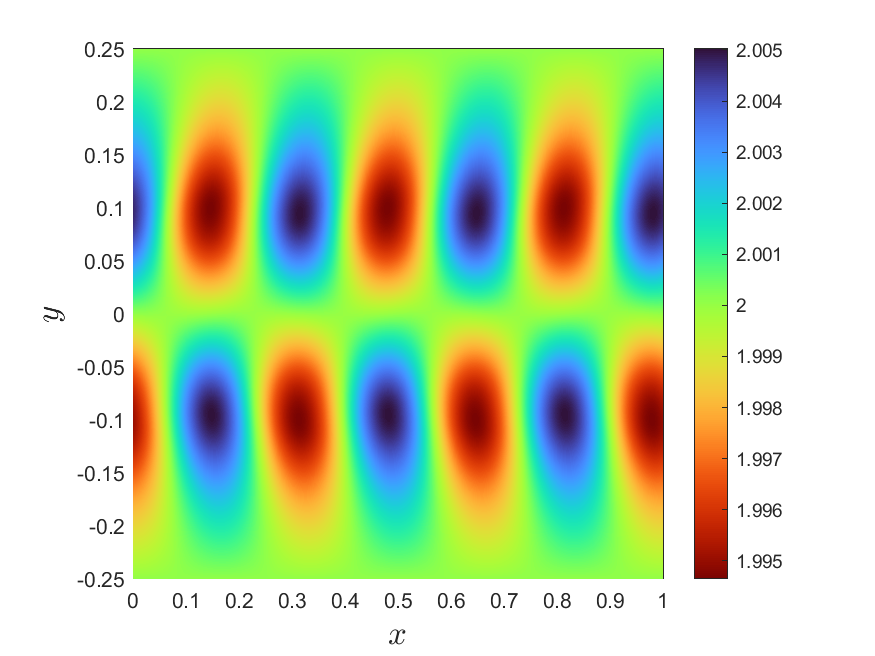}\label{fig:INSTDJ_p}}
    \caption{Instantaneous visualization of the inviscid doubly periodic jet flow at $t/t_c=4$. Results have been obtained on a refined grid consisting in $N_x\times N_y=257\times 129$ evenly spaced points to guarantee a satisfactory visualization of the main flow-field structures. Here, CFL has been set to $0.5$.}
    \label{fig:INSTDJ}
\end{figure}

Fig.~\ref{fig:INSTDJ} reports instantaneous visualizations of the entropy and pressure fields for illustrative purposes, carried out with the AEC-TP$^{(5)}$ discretization at a higher spatial resolution ($257\times129$). Entropy is shown in Fig.~\ref{fig:INSTDJ_s} and it remarks the roll-up behaviour provided by the transversal jet. As expected, pressure variations (shown in Fig.~\ref{fig:INSTDJ_p}, appearing as compression/expansion zones arising from the rolled-up regions) are relatively small if compared to thermal ones, which represent the major contribution to the entropy field.

\begin{figure}[tb]
    \centering
     \subfloat[Asymptotic convergence on entropy conservation]{\includegraphics[width=0.5\linewidth]{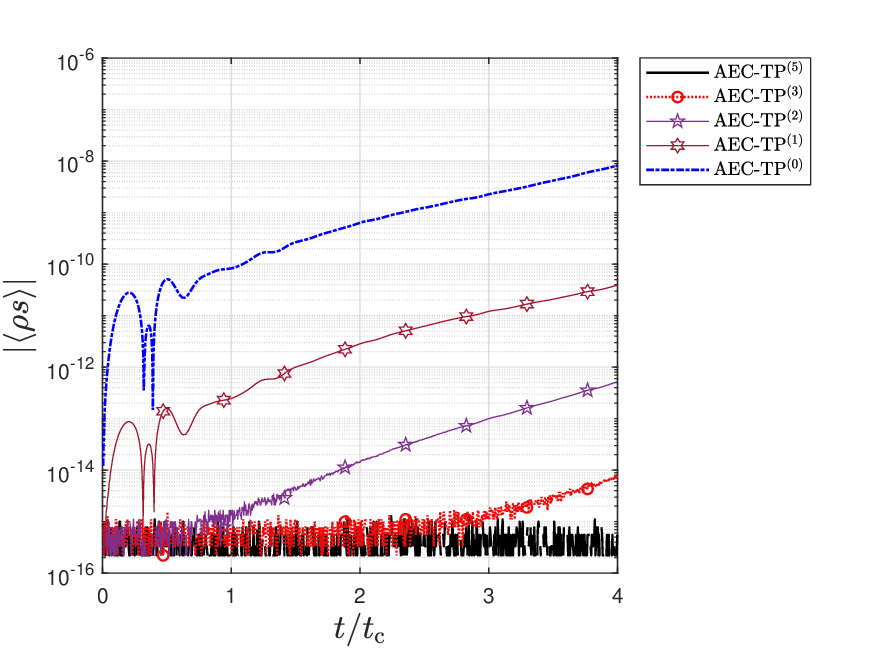}\label{fig:DJ_conv}}
     \subfloat[Entropy conservation]{\includegraphics[width=0.5\linewidth]{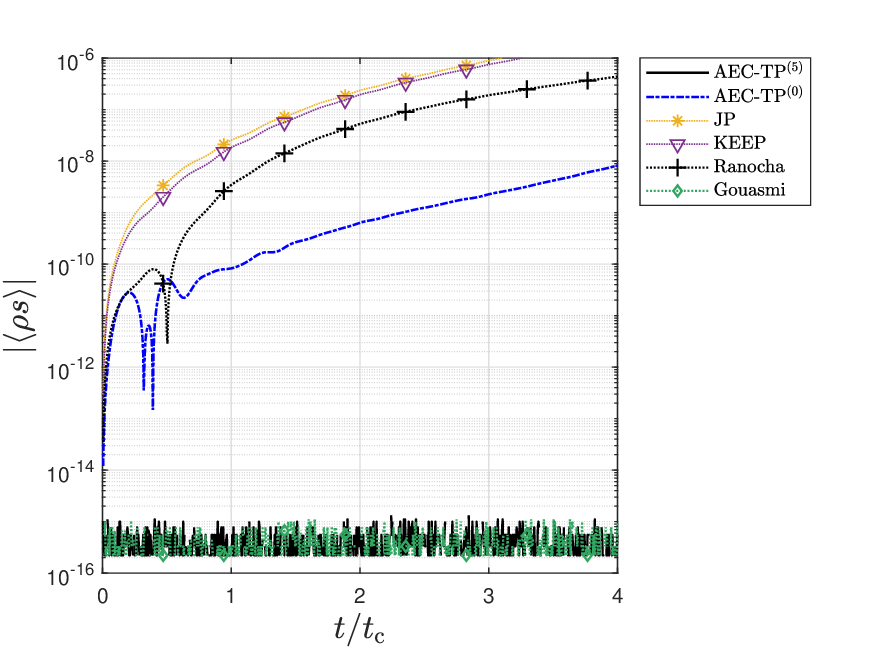}\label{fig:DJ_EC}}
    \caption{Inviscid doubly periodic jet flow. (a) asymptotic convergence of the novel AEC discretization. (b) error on entropy conservation.}
    \label{fig:DJ}
\end{figure}

Fig.~\ref{fig:DJ_conv} shows the asymptotic convergence of the AEC-TP$^{(N)}$ formulation with the increasing number of expansion terms.
For this particular test case, a number $N=5$ of terms in the asymptotic expansion is sufficient to reach machine-zero entropy production, indicating convergence to the exact EC-TP scheme in Eq.~\eqref{eq:ECFlux_exact}.
In Fig.~\ref{fig:DJ_EC} the error on entropy conservation is shown for the various numerical schemes here considered. 
Both the AEC-TP$^{(5)}$ and the fluxes proposed by \citeauthor{Gouasmi_CMAME_2020} show exact entropy conservation, while the discretizations designed for ideal gases reveal an error which accumulates in time. The same applies to the AEC-TP$^{(0)}$ which, however, shows a better behaviour for this particular test case. 

\subsection{Inviscid three-dimensional Taylor--Green Vortex}\label{Sec:Results_TGV}
As a second test case, we consider a three-dimensional turbulent flow, specifically the inviscid, periodic Taylor--Green Vortex (TGV). The main reason behind the choice of such configuration lies not only in the assessment of the EC properties, which clearly emerge in the previous test case as well as in \cite{Aiello_JCP_2025}, but also in the understanding of the effects of the discretization of the pressure term of the momentum equation, which differentiates the proposed formulation from other existing EC discretizations for thermally perfect gases \cite{Gouasmi_CMAME_2020,Oblapenko_C&F_2025}. We will show that the discretization of the pressure term as $\overline{p}$ instead of $R\overline{\rho}/(\overline{1/T})$ produces significant differences in terms of turbulent fluctuations and kinetic-energy evolution.\par
The TGV is solved in the tri-periodic box $[0,2\pi L]^3$, with $L=1$, discretized in $N_p=32^3$ evenly spaced points by means of sixth-order spatial discretizations and the initial conditions are
\begin{equation*}
    \begin{dcases}
        \rho(x,y,z,0) & = \rho_0\\
        u(x,y,z,0) & = u_0\sin{x}\cos{y}\cos{z}\\
        v(x,y,z,0) &  = -u_0\cos{x}\sin{y}\cos{z}\\
        w(x,y,z,0) & = 0\\
        p(x,y,z,0) & = p_0 + \frac{\rho_0 u_0^2}{16}\left(\cos{2x} + \cos{2y}\right)\left(\cos{2z} + 2\right)
    \end{dcases}
\end{equation*}
with $\rho_0= 1$ and $p_0=2.5$. All the schemes we are testing have been implemented in the open-source, GPU-accelerated code STREAmS-2.0~\cite{Bernardini_CPC_2023} and the dimensionless setting there used has been retained in our tests.
The Mach number---which defines the value of $u_0$---is set as $\Ma=0.1$ to hold the flow nearly incompressible.
The CFL number is set to $0.1$, as it has been observed to be sufficiently small to consider time-integration errors negligible, and the simulation is carried out for $100$ characteristic times, where one characteristic time is defined as $t_{\mathrm{c}}=L/u_0$.
\begin{figure}[tb]
    \centering
    \subfloat[Entropy conservation]{\includegraphics[width=0.5\linewidth]{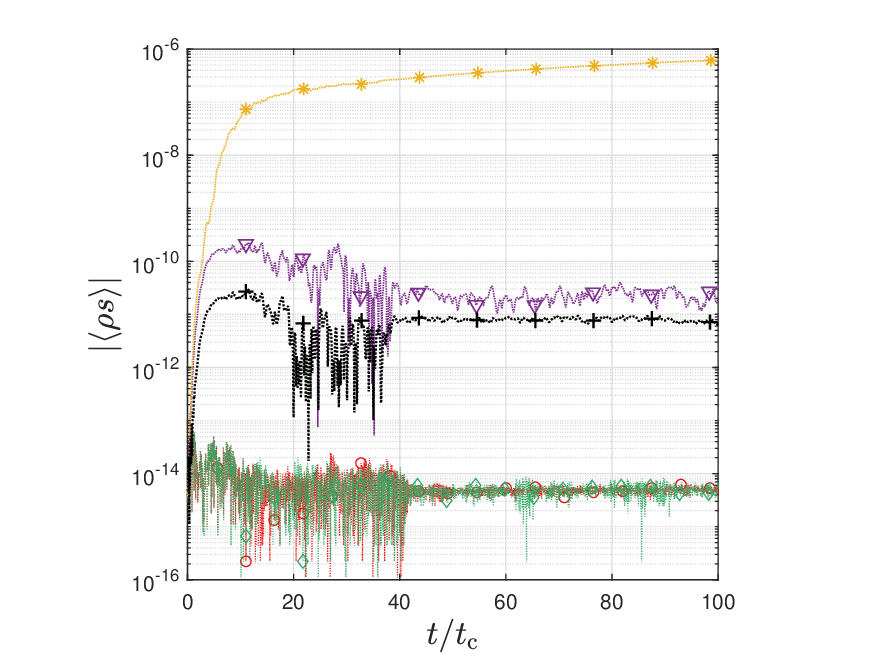}\label{fig:TGV_s}}
    \subfloat[Kinetic energy preservation]{\includegraphics[width=0.5\linewidth]{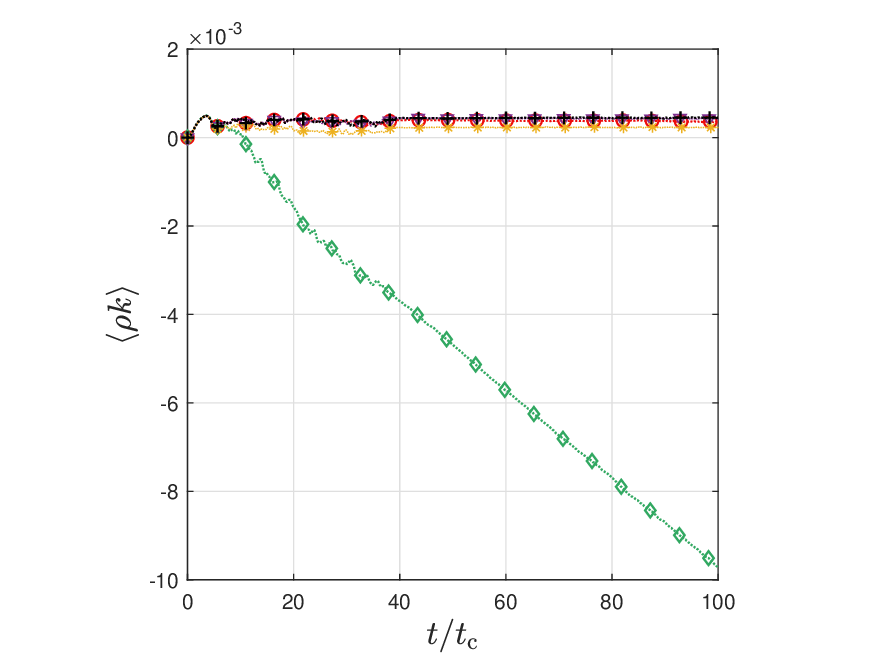}\label{fig:TGV_kin}}\\
     \subfloat[Internal energy preservation]{\includegraphics[width=0.5\linewidth]{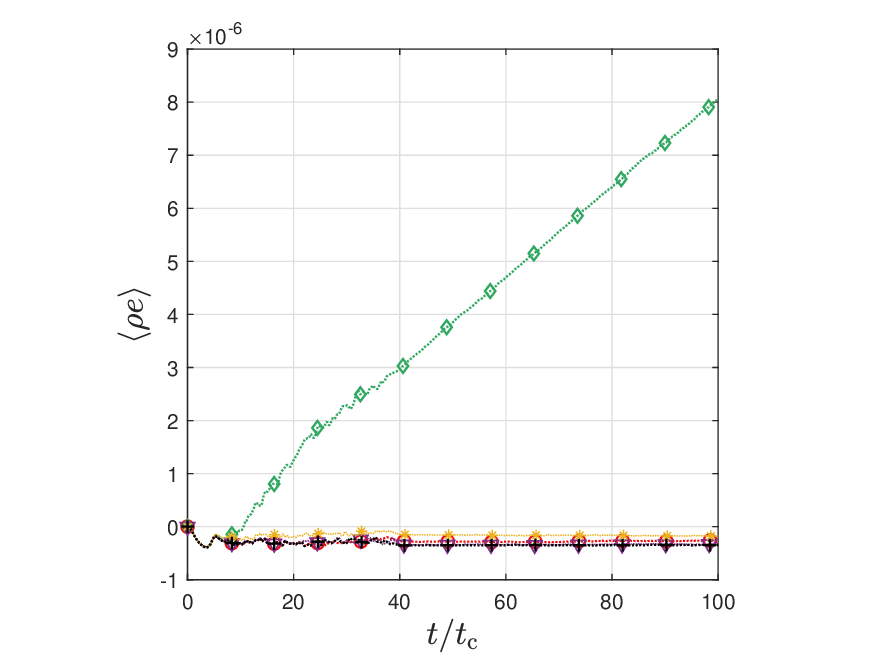}\label{fig:TGV_eint}}
     \subfloat[Kinetic energy-related pressure term evolution]{\includegraphics[width=0.5\linewidth]{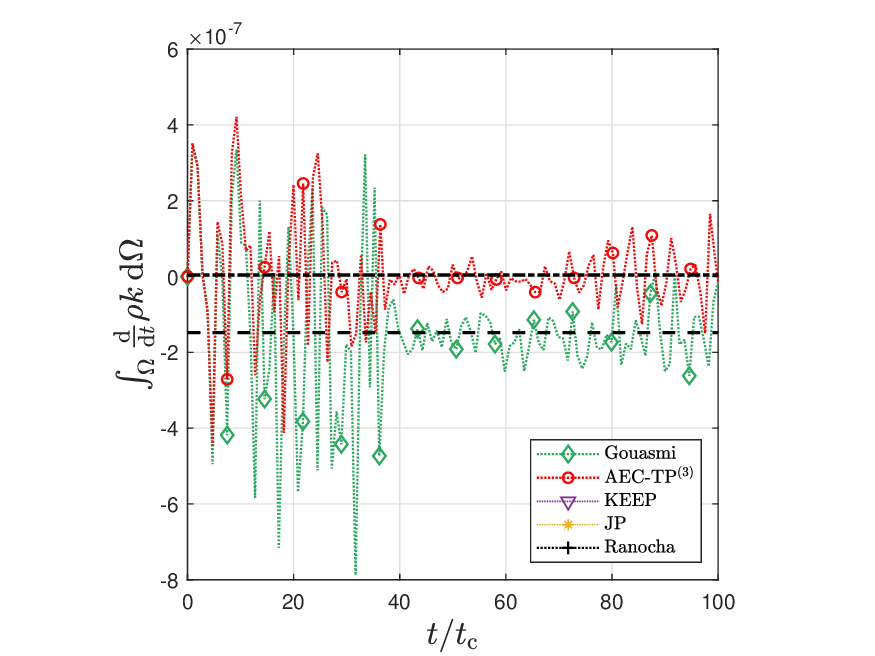}\label{fig:TGV_udp}}
      \caption{Inviscid three-dimensional Taylor--Green vortex at $\Ma=0.1$. (a) entropy-conservation error (b) kinetic energy evolution (c) internal energy evolution (d) kinetic energy-related pressure term evolution. For graphical clarity, data are sampled every 30 time steps for Figs.~(a)-(c) and every 500 time steps for Fig.~(d). 
      }
    \label{fig:TGV1}
\end{figure}

\begin{figure}
    
    \subfloat[Temperature fluctuations]{\includegraphics[width=0.5\linewidth]{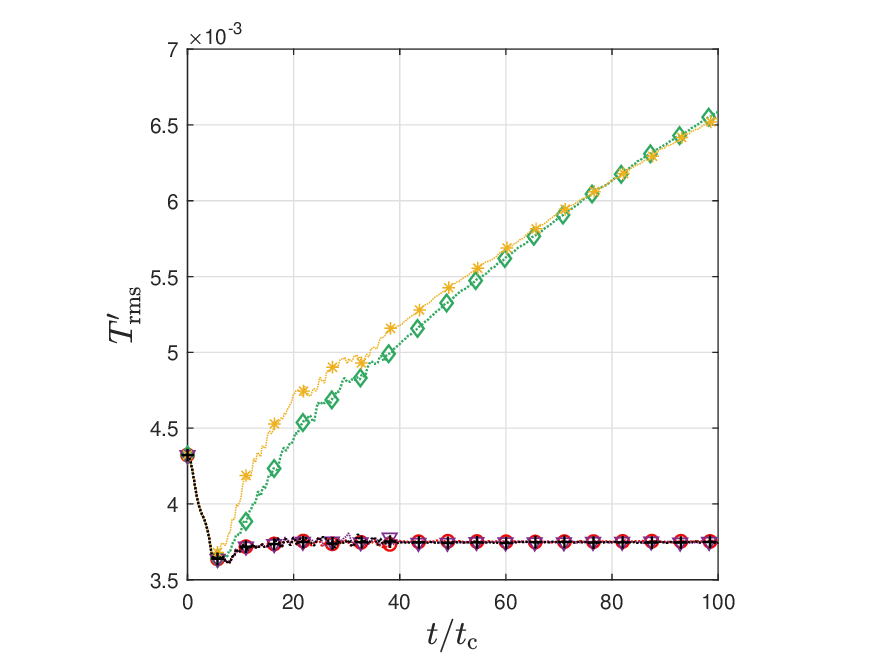}\label{fig:TGV_rho_fluct}}
    \subfloat[Density fluctuations]{\includegraphics[width=0.5\linewidth]{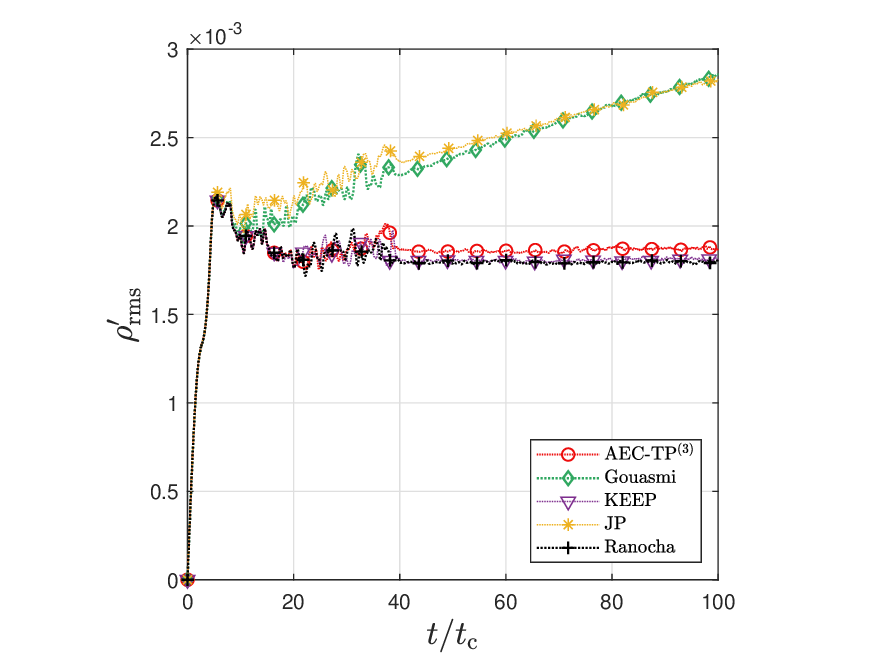}\label{fig:TGV_T_fluct}}
    \caption{Inviscid three-dimensional Taylor--Green vortex at $\Ma=0.1$. (a) r.m.s. temperature fluctuations. (b) r.m.s. density fluctuations. For graphical clarity, data are sampled every 30 time steps.
    }
    \label{fig:TGV}
\end{figure}
Fig.~\ref{fig:TGV_s} shows the entropy-conservation properties of the considered schemes. As already observed in~\cite{Aiello_JCP_2025}, the TGV produces ‘‘well-behaved" entropy evolutions regardless of the gas model if the numerical scheme is, even approximately, EC for ideal gases.
We could attribute such behaviour to the test case itself, which provides relatively small variations of the thermodynamic quantities. 
However, exact conservation of entropy for the AEC-TP$^{(N)}$ with $N=3$ (which is the value guaranteeing convergence for this test case) and for the Gouasmi scheme is evident in Fig.~\ref{fig:TGV_s}, where the nondimensionalized error on entropy conservation is zero to machine precision.
The KEEP and Ranocha formulations display entropy production which is approximately three orders of magnitude greater, whereas the JP scheme completely fails to keep the entropy production under control.
Despite their almost identical behaviour on entropy conservation, the AEC-TP and Gouasmi formulations show 
important discrepancies when evaluated on other observables.  An example is given by the evolution of global kinetic energy, which is reported in  Fig.~\ref{fig:TGV_kin}.
Although total kinetic energy is not conserved in general in compressible inviscid flows, for the inviscid TGV test under consideration it is expected that global kinetic energy does not significantly vary in time~\cite{Gassner_JCP_2016}.
The Gouasmi formulation is designed to be KEP, as all the other formulations we are considering, in the sense that the discretization of the advective terms does not influence the kinetic energy balance. However, as documented for the Chandrashekar formulation in~\cite{Gassner_JCP_2016}, the discretization of the pressure term of the momentum equation as $R\mean{\rho}/(\mean{1/T})$ produces nonnegligible kinetic-energy dissipation with respect to all of the other schemes, which enforce the discretization of the pressure flux as $\mean{p}$. 
Fig.~\ref{fig:TGV_eint} shows the evolution of global internal energy, with an analogous production for the Gouasmi scheme due to the transfer of energy allowed by the pressure-work terms.
The effect of the spatially integrated pressure work contribution is shown in Fig.~\ref{fig:TGV_udp}, where the integrated time derivative of kinetic energy is shown as representative of the nonconservative pressure term in the discrete kinetic energy evolution equation, since the convective term is in locally conservative form. The nonzero average displayed by the Gouasmi scheme reflects its failure in keeping the global kinetic energy constant.

An additional difference between the novel and previous formulations is seen in the fluctuations of the thermodynamic quantities of the density $\rho'_{\mathrm{rms}}$ and temperature $T'_{\mathrm{rms}}$, reported in Fig.~\ref{fig:TGV}, in
which $\phi'_{\mathrm{rms}} = \sqrt{\frac{1}{N_p}\sum_{i=1}^N (\phi_i - \phi_\mu)^2}$ and $\phi_\mu = \frac{1}{N_p} \sum_{i=1}^{N_p} \phi_i$ is the average value of $\phi$ in the domain.
In fact, as it has been already shown for the JP scheme (\cite{Pirozzoli_JCP_2010,DeMichele_C&F_2023}), it seems that energy fluxes whose pressure term is not built as in~\eqref{eq:General_Flux_Form} (i.e.~JP and Gouasmi formulations) produce fluctuations which do not reach a steady-state after the thermalization of the flow field, which is the expected result for homogeneous isotropic turbulence calculations~\cite{Honein_JCP_2004}.

\subsection{Sod shock tube test}
\label{sec:Sod_TP}

The EC schemes derived in the previous sections are constructed under the assumption of smooth solutions. However, compressible flows of practical interest typically involve discontinuities such as shocks and contact surfaces. In these regimes, purely non-dissipative discretizations are not appropriate: although they preserve a discrete entropy balance by construction, they do not enforce the entropy inequality required by the second law and generally produce spurious oscillations in under-resolved shocked regions.

A common strategy to address this issue is to employ a shock sensor and locally switch to a more dissipative discretization, such as WENO. This approach is followed, for instance, in~\cite{Bernardini_CPC_2023}, and entropy-stable (ES) variants of WENO have also been developed~\cite{Fisher_JCP_2013b}. In the present work, we adopt an alternative and widely used strategy: rather than switching schemes, we augment the EC flux with a suitable dissipation term so as to obtain an ES formulation.

To recover physically admissible solutions, the semi-discrete scheme must satisfy a discrete entropy inequality instead of an exact entropy conservation property. This can be achieved by adding a consistent dissipative contribution to the EC numerical flux. Here, we consider a Rusanov (local Lax--Friedrichs, LLF) regularization, leading to the ES flux
\begin{equation*}
    \mF^{ES}(\mathbf{u}_i,\mathbf{u}_{i+1})
    =
    \mF^{EC}(\mathbf{u}_i,\mathbf{u}_{i+1})
    -
    \frac{1}{2}\,\lambda_{\max}
    \left(
        \mathbf{u}_{i+1} - \mathbf{u}_i
    \right),
\end{equation*}
where $\mF^{EC}$ denotes the baseline entropy-conservative flux and $\lambda_{\max}$ is the maximum local signal speed,
\begin{equation*}
    \lambda_{\max} = \max \left( |u_i| + c_i,\; |u_{i+1}| + c_{i+1} \right),
\end{equation*}
with $c$ the thermodynamic speed of sound of the TP gas model. This dissipation yields an ES scheme in the sense that the resulting formulation satisfies the correct entropy inequality at the semi-discrete level (see, e.g.,~\cite{Ranocha_JSC_2018}). 
\begin{figure}
    \centering
    \subfloat[Velocity]{\includegraphics[width=0.5\linewidth]{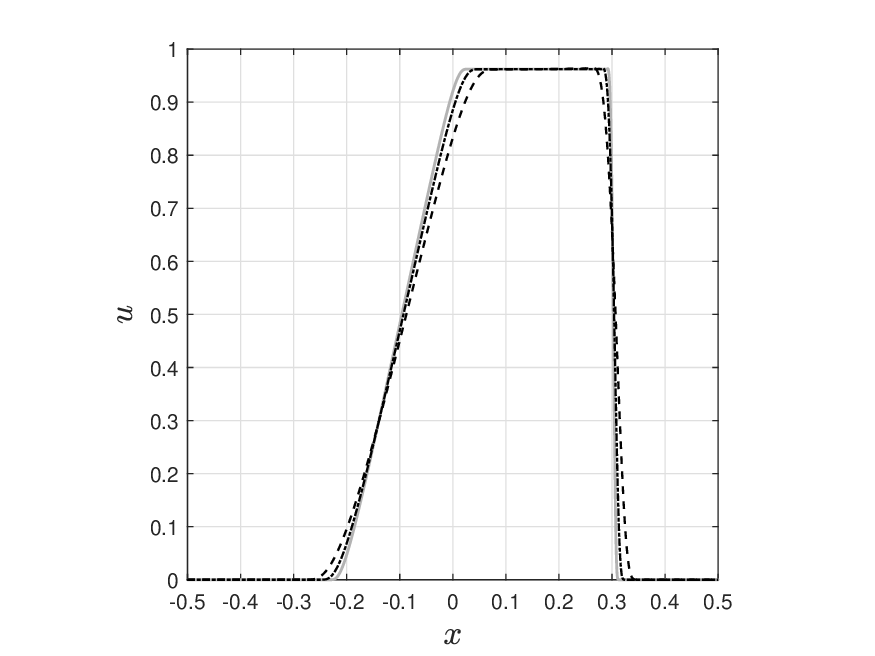}}
    \subfloat[Pressure]{\includegraphics[width=0.5\linewidth]{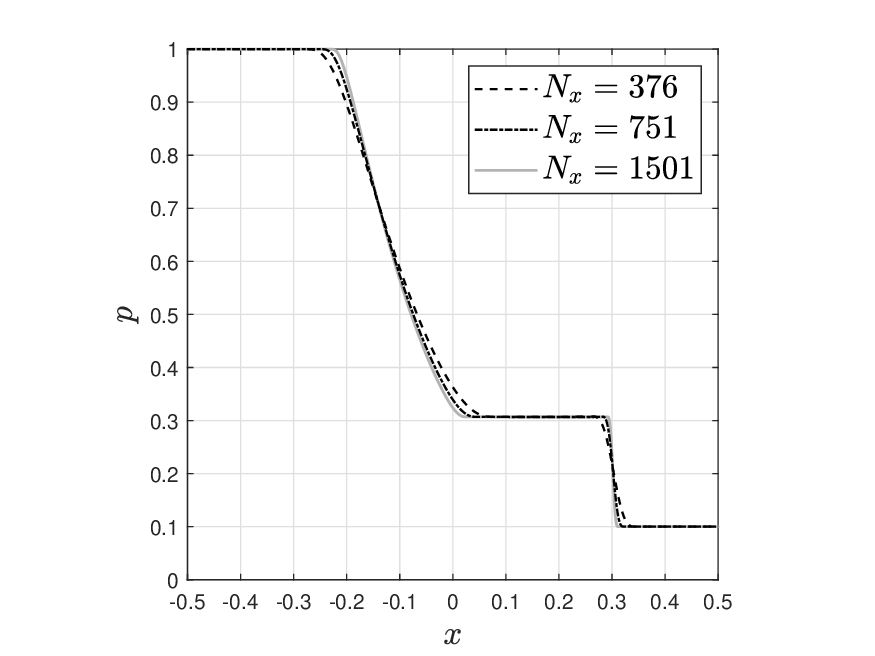}}\\
    \subfloat[Density]{\includegraphics[width=0.5\linewidth]{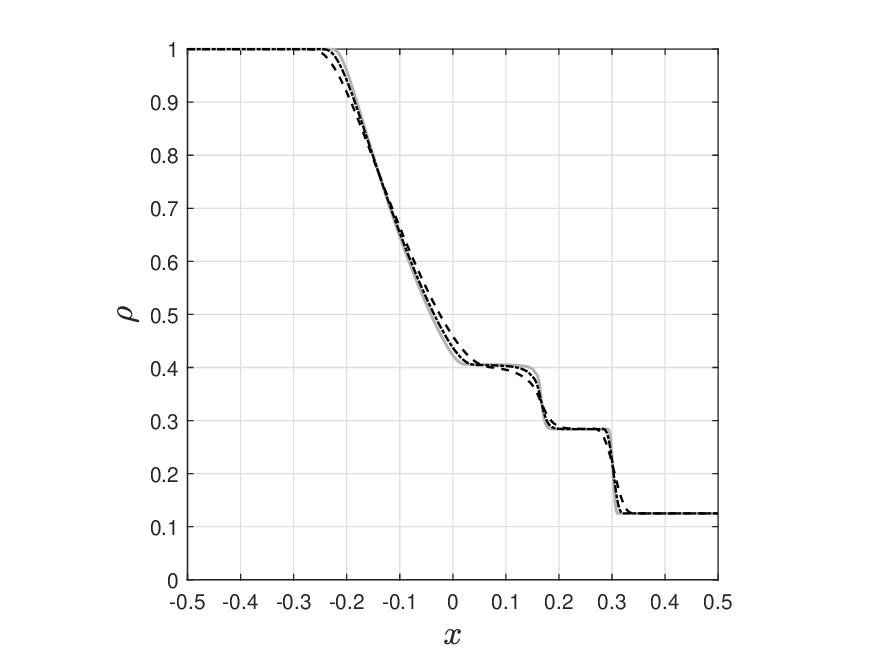}}
    \subfloat[Volumetric entropy integral]{\includegraphics[width=0.5\linewidth]{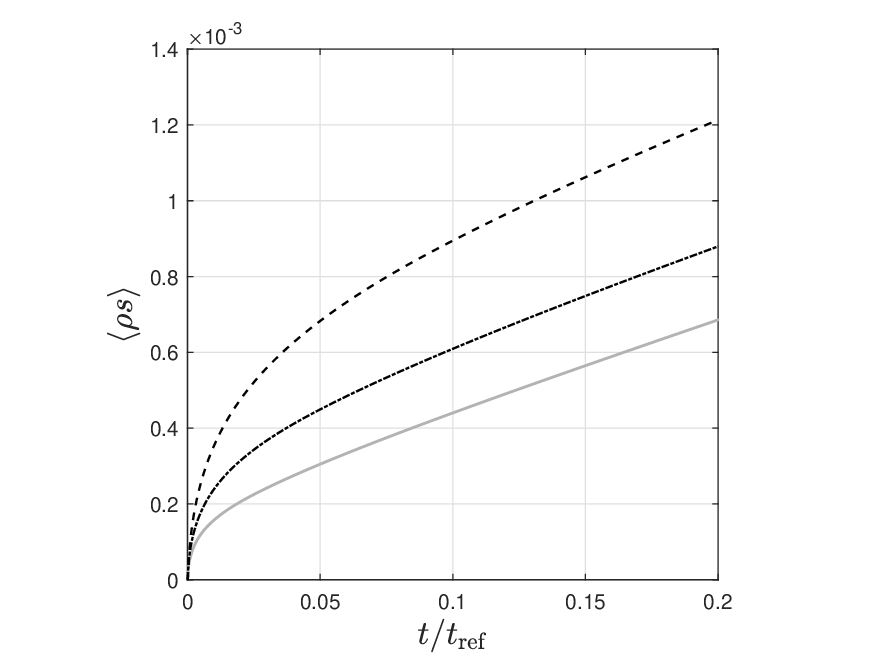}}
    \caption{One-dimensional Sod shock tube problem. Results obtained with the EC-TP scheme augmented with a LLF dissipation term for different numbers of spatial discretization points $N_x$.}
    \label{fig:SOD}
\end{figure}
Although the LLF term is robust and straightforward to implement, it may introduce excessive diffusion, in particular across contact discontinuities. To mitigate this effect, we modulate the dissipative term with a simple pressure-jump sensor~\cite{Chandrashekar_CCP_2013},
\begin{equation*}
    \Xi = \sqrt{\frac{|p_{i+1} - p_i|}{p_{i+1} + p_i}},
\end{equation*}
which scales the LLF contribution. The coefficient $\Xi$ is close to unity in the presence of strong pressure gradients (e.g.~shocks) and small in smooth regions, thereby reducing unnecessary dissipation away from discontinuities. The resulting flux can be written as
\begin{equation*}
    \mF^{ES}
    =
    \mF^{EC}
    -
    \frac{1}{2}\,\Xi\,\lambda_{\max}
    \left(
        \mathbf{u}_{i+1} - \mathbf{u}_i
    \right).
\end{equation*}

To assess the performance of this ES extension in the presence of shocks, we consider the classical Sod shock tube problem. The one-dimensional computational domain $\Omega=[-1,1]$ is divided into two constant states by a diaphragm located at its midpoint, $x=0$. The initial left and right states are
\begin{equation*}
\begin{dcases}
    (\rho,u,p) =(1,0,1)\quad &\mathrm{for}\;\,x<0,\\
    (\rho,u,p) = (0.125,0,0.1)\quad &\mathrm{for}\;\,x\geq0.
\end{dcases}
\end{equation*}
At $t=0$ the diaphragm is removed, generating a right-moving shock, a left-moving rarefaction wave, and a contact discontinuity. The simulation is carried out up to $t/t_{\mathrm{ref}}=0.2$, where $t_{\mathrm{ref}}=L_x/c_{\mathrm{max}}$, with $c_{\mathrm{max}}=\max_\Omega(\sqrt{\gamma(T)RT})$ and $L_x=1$. CFL is equal to $0.1$, and it is based on the reference velocity $c_{\mathrm{max}}$.

Fig.~\ref{fig:SOD} shows results for different numbers of discretization points $N_x=\{376,751,1501\}$ contained in the interval $[-1/2,1/2]$.
Regarding spatial discretization, simulations have been carried out with the EC-TP formulation (Eqs.~\eqref{eq:General_Flux_Form}, \eqref{eq:ECFlux_tp} and \eqref{eq:ECFlux_exact}) for the smooth regions in its second-order accuracy version. 
The proposed strategy furnished excellent solutions which converge with increasing grid points. No visible oscillations are observed and the physical discontinuities are clearly highlighted.

\section{Conclusions}
\label{sec:Conclusions}
In this paper, we have developed a spatial discretization of the compressible Euler equations that guarantees exact discrete conservation of entropy in the case of thermally perfect gases and we have validated it on both 2D and 3D test cases. 
In addition to being EC and conservative of the primary invariants, the scheme is also preservative of kinetic energy and can be effortlessly extended to higher-order formulations.
The theory presented here is also shown to be applicable in the case of multicomponent gas mixtures.

Starting from the EC condition valid for a generic real gas, we demonstrated how the specific structure of the EoS for perfect gases can be exploited to eliminate singularities in the formulation of the numerical flux, thereby simplifying the resulting expressions.
The only remaining singularity arises from the presence of logarithmic means, which can be addressed using locally the algorithm by \citet{Ismail_JCP_2009} or by constructing a hierarchy of asymptotically entropy-conserving schemes thorough a Taylor series expansion, as in the case of calorically perfect gases.
Importantly, this result is independent of the specific thermally perfect model used to relate internal energy and absolute temperature, supporting its use in a wide range of applications.
As illustrative examples, we have considered both a polynomial-based model for the isochoric specific heat and the Rigid-Rotor Harmonic-Oscillator model, each leading to a different variant of the EC scheme.

In comparing the proposed scheme to existing EC schemes, the key distinction lies in the treatment of the pressure terms in the momentum and total energy equations.
The difference proves critical, since our approach yields a more accurate evolution of kinetic energy and better control over fluctuation-related statistical quantities.
This improvement stems from the methodology employed: rather than relying on Tadmor’s approach, we adopted a more direct and flexible strategy that facilitates the incorporation of thermodynamic consistency, which is especially valuable for complex or realistic gas models.

Numerical experiments further support the theoretical predictions.
Simulations of inviscid doubly periodic jet flows confirm the entropy-conserving properties of the proposed scheme and that the AEC-TP hierarchy is able to recover exact entropy conservation when a sufficient number of terms is retained. 
The results from the 3D Taylor--Green vortex highlight clear advantages over the existing EC schemes in capturing both the evolution of kinetic energy and small-scale fluctuations.
These findings underline the potential of the proposed EC schemes as a robust and accurate tool for simulating compressible flows involving thermally perfect gases.

The results presented in this paper open several avenues for future development.
For instance, further extensions could focus on deriving entropy-stable discretizations for the viscous terms as well, guaranteeing the correct entropy monotonicity, thus obtaining fully entropy-consistent spatial discretizations for arbitrary equations of state. Moreover, while the current formulation is restricted to nonreacting gas mixtures, the generality of the approach suggests potential extensions to reacting flows, including chemically reactive mixtures and multitemperature models.

\section*{Acknowledgments}
We acknowledge the CINECA award under the ISCRA initiative, for the availability of high-performance computing resources and support.
We thank Dr.~Ayaboe K.~Edoh and Marco Artiano for reading an early draft of the manuscript and providing thoughtful feedback.

\bibliographystyle{model1-num-names}
\bibliography{Biblio_KEP_Compr}

\end{document}